\documentclass{article}

\PassOptionsToPackage{sort&compress,numbers,super}{natbib}
\usepackage[preprint]{neurips_2023}

\usepackage[utf8]{inputenc} 
\usepackage[T1]{fontenc}    
\usepackage{hyperref}       
\usepackage{url}            
\usepackage{booktabs}       
\usepackage{amsfonts}       
\usepackage{nicefrac}       
\usepackage{microtype}      
\usepackage{xcolor}         
\usepackage[ruled,vlined]{algorithm2e}      
\usepackage[numbers,sort&compress]{natbib}
\usepackage{graphicx}

\title{Augmented Memory: Capitalizing on Experience 
Replay to Accelerate \textit{De Novo} Molecular Design}

\author{
  Jeff Guo\textsuperscript{1,2}, Philippe Schwaller\textsuperscript{1,2} \\
  \textsuperscript{1}Laboratory of Artificial Chemical Intelligence (LIAC), Institut des Sciences et Ing\'{e}nierie Chimiques, \\ Ecole Polytechnique F\'{e}d\'{e}rale de Lausanne (EPFL), Lausanne, Switzerland \\
  \textsuperscript{2}National Centre of Competence in Research (NCCR) Catalysis, \\ Ecole Polytechnique F\'{e}d\'{e}rale de Lausanne (EPFL), Lausanne, Switzerland \\
  \texttt{\{jeff.guo,philippe.schwaller\}@epfl.ch} \\
}

\begin{document}

\maketitle
\begin{abstract}
  Sample efficiency is a fundamental challenge in \textit{de novo} molecular design. Ideally, molecular generative models should learn to satisfy a desired objective under minimal oracle evaluations (computational prediction or wet-lab experiment). This problem becomes more apparent when using oracles that can provide increased predictive accuracy but impose a significant cost. Consequently, these oracles cannot be directly optimized under a practical budget. Molecular generative models have shown remarkable sample efficiency when coupled with reinforcement learning, as demonstrated in the Practical Molecular Optimization (PMO) benchmark. Here, we propose a novel algorithm called Augmented Memory that combines data augmentation with experience replay. We show that scores obtained from oracle calls can be reused to update the model multiple times. We compare Augmented Memory to previously proposed algorithms and show significantly enhanced sample efficiency in an exploitation task and a drug discovery case study requiring both exploration and exploitation. Our method achieves a new state-of-the-art in the PMO benchmark which enforces a computational budget, outperforming the previous best performing method on 19/23 tasks. The code is available at \url{https://github.com/schwallergroup/augmented_memory}.
\end{abstract}

\section{Introduction}

A quintessential task in any molecular discovery campaign is identifying promising candidate molecules amidst an enormous chemical space~\cite{sanchez-lengeling_inverse_2018}. With the democratization of computing resources, computational oracles can be deployed to query larger chemical spaces in search of the desired property profile. The use of such oracles has enabled researchers to identify functional materials~\cite{westermayr_high-throughput_2023}, therapeutics~\cite{lyu2019ultra, zhavoronkov_deep_2019, ren_alphafold_2023}, and catalysts~\cite{seumer2022computational}, thus accelerating chemical discovery. However, there is generally a trade-off between oracle predictive accuracy and inference cost, such that the computational budget imposes a pragmatic constraint. Provided a sufficiently sample efficient model, it is conceivable for wet-lab experiments to be the oracle itself, as enabled by a high-throughput experimentation platform. Correspondingly, designing computational workflows and algorithms that are performant under minimal oracle calls is widely beneficial to the field of molecular design.

Recent advancements in \textit{de novo} molecular design have positioned generative methods as a complementary approach to traditional virtual screening~\cite{lyu2019ultra, sadybekov_synthon-based_2022}. Core advantages of these models include the ability to sample chemical space outside the training data and by coupling an optimization algorithm, goal-directed learning can be achieved~\cite{meyers_novo_2021}. Although the field is relatively nascent, molecular generative models have identified experimentally validated therapeutic molecules~\cite{yoshimori_design_2021, zhavoronkov_deep_2019, ren_alphafold_2023, korshunova_generative_2022} and organocatalysts~\cite{seumer2022computational}. An important shared commonality between these success stories is the inclusion of relatively computationally expensive oracles that are optimized. In drug design, molecular docking is frequently used while in catalyst and materials design, quantum mechanical properties are of interest. Correspondingly, many generative models proposed in recent years have competed to demonstrate accelerated optimization of these properties. However, the heterogeneity of the assessment protocols makes comparisons difficult. Recently, Gao et al.~\cite{gao_sample_2022} propose the Practical Molecular Optimization (PMO) benchmark which assesses 25 molecular generative models across 23 tasks, enforcing a computational budget of 10,000 oracle calls. Their results show that REINVENT~\cite{olivecrona_molecular_2017, blaschke_reinvent_2020}, a recurrent neural network (RNN)-based generative model operating on simplified molecular-input line-entry system (SMILES)~\cite{weininger_smiles_1988} is, on average, the most sample efficient generative model. REINVENT~\cite{olivecrona_molecular_2017, blaschke_reinvent_2020} uses a policy-based reinforcement learning (RL) algorithm to optimize a reward function in a goal-directed approach. Recently, alternative algorithms have been proposed in the form of Best Agent Reminder (BAR)~\cite{atance2022novo} and Augmented Hill Climbing (AHC)~\cite{thomas_augmented_2022} which both introduce bias towards high rewarding molecules to improve sample efficiency. Other studies show that experience replay, where the highest rewarding molecules sampled are stored and replayed to the model, improves sample efficiency.~\cite{blaschke_reinvent_2020, korshunova_generative_2022} More recently, Bjerrum et al.~\cite{bjerrum_faster_2023} proposed Double Loop RL to take advantage of the non-injective nature of SMILES and the ease with which they can be augmented. By obtaining different SMILES sequences for the same molecule, oracle scores can be re-used to perform multiple updates to the Agent. Their results show accelerated learning while maintaining the diversity of results, an aspect missing in many proposed benchmarks. 

Sample efficiency is a limiting factor to enabling more exploration of chemical spaces of interest, such as in drug discovery where the reward is sparse, i.e., finding a needle in the haystack. In this paper, we highlight the importance of experience replay in policy-based RL algorithms for molecular generation. We propose a novel algorithm called Augmented Memory that combines experience replay with SMILES augmentation. We further propose Selective Memory Purge which removes entries in the replay buffer with undesired chemical scaffolds and jointly address sample efficiency and diversity. The main contributions of this paper are:

\begin{itemize}
    \item We explicitly highlight the importance of experience replay on the sample efficiency of REINVENT and all proposed algorithmic modifications.
    \item We propose a novel algorithm called Augmented Memory which significantly outperforms all previous algorithms in sample efficiency. This is demonstrated in an exploitation task and a drug discovery case study.
    \item We propose a method called Selective Memory Purge, which can be used in conjunction with Augmented Memory to generate diverse molecules while retaining enhanced sample efficiency.
    \item We expand the PMO benchmark~\cite{gao_sample_2022} by adding Augmented Memory and BAR~\cite{atance2022novo} implementations. We further add experience replay to the implemented version of AHC\cite{thomas_augmented_2022, thomas2022re} for comparison. Our algorithm achieves a new state-of-the-art and outperforms the previous state-of-the-art, REINVENT~\cite{olivecrona_molecular_2017, blaschke_reinvent_2020}, on 19/23 tasks.
\end{itemize}

\section{Related Work}

\textbf{Goal-directed Molecular Design with Policy-based Reinforcement Learning.} Molecular generation can be framed as a policy-based RL problem, where a base model (Prior) is trained on a general dataset and fine-tuned (Agent) to generate molecules with desired property profiles. Existing works that follow this paradigm include SMILES-based RNNs~\cite{olivecrona_molecular_2017, popova_deep_2018, blaschke_reinvent_2020, segler2018generating}, generative adversarial networks (GANs)~\cite{goodfellow_generative_2014, sanchez2017optimizing, putin_reinforced_2018, guimaraes_objective-reinforced_2018, de_cao_molgan_2022}, variational autoencoders (VAEs)~\cite{kingma_auto-encoding_2022, zhavoronkov_deep_2019}, graph-based models~\cite{you_graph_2019, jin_multi-objective_2020,mercado_graph_2021, atance2022novo}, and GFlowNets~\cite{bengio2021flow}. While all methods can generate valid molecules and the policy can be fine-tuned via RL, none of the previous methods jointly address sample efficiency and a reliable mechanism to mitigate mode collapse. We note that GFlowNets~\cite{bengio2021flow} by construction can achieve diverse sampling but are not as sample efficient as demonstrated in the PMO benchmark~\cite{gao_sample_2022}. By contrast, SMILES-based models, particularly REINVENT~\cite{olivecrona_molecular_2017, blaschke_reinvent_2020}, have been shown to be amongst the most sample efficient molecular generative models, even when compared to the newest proposed models~\cite{gao_sample_2022}. Moreover, their ability to learn complex molecular distributions~\cite{flam2022language} and satisfy multi-parameter optimization (MPO) objectives has been shown in diverse benchmarks, such as GuacaMol~\cite{brown_guacamol_2019}, MOSES~\cite{polykovskiy_molecular_2020}, and PMO~\cite{gao_sample_2022}. Our proposed Augmented Memory algorithm builds on this observation and exploits the non-injective nature of SMILES.

\textbf{Sample Efficiency in Molecular Design.} Many existing policy-based RL works for molecular design are based on the REINFORCE~\cite{williams_simple_1992} algorithm, particularly for models operating on SMILES. Algorithmic alternatives present a unifying theme of using biased gradients to direct the policy towards chemical space with high reward. Neil et al.~\cite{neil_exploring_2018} explored Hill Climbing (HC) and Proximal Policy Optimization (PPO)~\cite{neil_exploring_2018}. Similarly, Atance et al. introduced Best Agent Reminder (BAR)~\cite{atance2022novo} which keeps track of the best agent and reminds the current policy of favorable actions. Thomas et al. introduced Augmented Hill Climbing (AHC)~\cite{thomas_augmented_2022}, a hybrid of HC and REINVENT's algorithm, which updates the policy at every epoch using only the top-k generated molecules and shows improved sample efficiency. However, sample efficiency by itself is not sufficient for practical applications of molecular generative models as one should aim to generate diverse molecules that satisfy the objective function. To address this limitation, Bjerrum et al. built directly on REINVENT and introduced Double Loop RL~\cite{bjerrum_faster_2023}. By performing SMILES augmentation, the policy can be updated numerous times per oracle call. Their results showed improved sample efficiency compared to AHC, while maintaining diverse sampling.

\textbf{Experience Replay for Molecular Design.} Experience replay was first proposed by Lin et al.~\cite{lin_self-improving_nodate} as a mechanism to replay past experiences to the model so that it can learn from the same experience numerous times. Two paradigms in RL are on-policy and off-policy where the model's actions are dictated by its current policy or a separate policy known as the behavior policy, respectively~\cite{fedus_revisiting_2020}. Experience replay is usually applied in off-policy methods as past experiences are less likely to be applicable to the current policy. In molecular design, experience replay has been proposed by Blaschke et al.~\cite{blaschke_reinvent_2020, blaschke_memory-assisted_2020} and Korshunova et al.~\cite{korshunova_generative_2022} to keep track of the best molecules sampled so far, based on their corresponding reward. Notably, both applications of experience replay are for on-policy learning using the REINFORCE\cite{williams_simple_1992} algorithm and only Korshunova et al. empirically show its benefit in sparse reward environments.  We note that a similar mechanism was proposed by Putin et al.~\cite{putin_reinforced_2018} using an external memory.

\section{Proposed Method: Augmented Memory}
\begin{figure}[ht!]
\centering
\includegraphics[width=\linewidth]{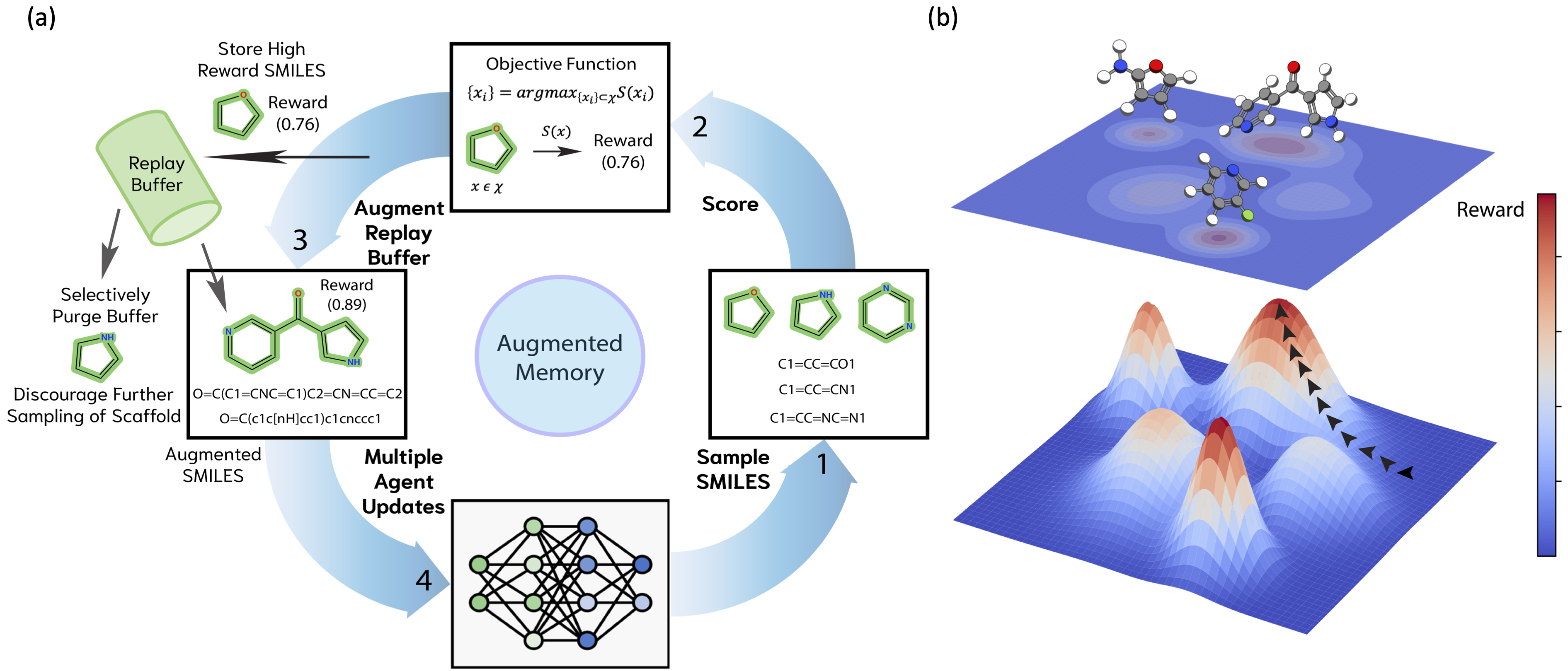} 
\caption{Augmented Memory. (a) The proposed method proceeds via four steps: 1. generate a batch of SMILES according to the current policy. 2. Compute the reward for the SMILES given the objective function. 3. Update the replay buffer to keep only the top $K$ molecules. Optionally, remove molecules from the replay buffer to discourage further sampling of specific scaffolds. Perform SMILES augmentation of both the sampled batch and the entire replay buffer. 4. Update the Agent and repeat step 3 $N$ times. (b) Schematic of the intended behavior. Augmenting the entire replay buffer and updating the Agent repeatedly directs chemical space exploration to areas of high reward.}
\label{fig:augmented}
\end{figure}

In this work, we extend the observations by Korshunova et al.~\cite{korshunova_generative_2022} and explicitly show the benefit of experience replay in dense reward environments, i.e., most molecules give at least \textit{some} reward, for on-policy learning given a static objective function. This static nature means that regardless of the current policy, high-rewarding molecules will always receive the same reward, which supports the efficacy of experience replay in the on-policy setting for molecular generation. Next, we combine elements of HC and SMILES augmentation with experience replay, and propose to update the policy at every fine-tuning epoch using the entire replay buffer. A reward shaping mechanism~\cite{wiewiora_reward_2010} is introduced by using these extremely biased gradients towards high rewarding chemical space which we show significantly improves sample efficiency. This section describes each component of Augmented Memory (Figure \ref{fig:augmented}) which is capable of performing MPO.

\textbf{Squared Difference Loss.} The molecular generative model builds directly on REINVENT~\cite{olivecrona_molecular_2017, blaschke_reinvent_2020} and is an autoregressive SMILES-based RNN using long short-term memory (LSTM)~\cite{hochreiter1997long} cells. The generative process is cast as an on-policy RL problem by defining the state space, $S_t$, and the action space, $A_t(s_t)$. Since REINVENT is a language model and samples tokens, $S_t$ denotes every intermediate sequence of tokens leading up to the fully constructed SMILES and $A_t(s_t)$ are the token sampling probabilities at every intermediate state. $A_t(s_t)$ is controlled by the policy, $\pi_{\theta}$, which is parameterized by the RNN. An assumption is that the SMILES generation process is Markovian (Equation \ref{eq:main-text-markovian}):

\begin{equation}
P(x) = \prod_{t=1}^{T} P(s_t \mid s_{t-1}, s_{t-2}, \ldots, s_1)
\label{eq:main-text-markovian}
\end{equation}

The Augmented Likelihood is defined as a linear combination between the Prior Likelihood and the scoring function, $S$, which returns a reward denoting the desirability of a given molecule and modulated by a hyperparameter sigma, $\sigma$ (Equation \ref{eq:main-text-augmented-likelihood}). The Prior Likelihood term acts to ensure the generated SMILES are syntactically valid, and has been shown to empirically enforce reasonable chemistry~\cite{olivecrona_molecular_2017, thomas_augmented_2022}.

\begin{equation}
\log\pi_{\theta_{\textsubscript{Augmented}}} = \log\pi_{\theta_{\textsubscript{Prior}}} + \sigma S(x)
\label{eq:main-text-augmented-likelihood}
\end{equation}

The policy is directly optimized by minimizing the squared difference between the Augmented Likelihood and the Agent Likelihood given a sampled batch, $B$, of SMILES constructed following the actions, $ a \in A^* $ (Equation \ref{eq:main-text-dap}):

\begin{equation}
L(\theta) = \frac{1}{|B|} \left[\sum_{a \in A^*}(\log\pi_{\theta_\textsubscript{Augmented}} - \log\pi_{\theta_\textsubscript{Agent}})\right]^2
\label{eq:main-text-dap}
\end{equation}

Taking the gradient of the loss function yields Equation \ref{eq:main-text-squared-loss-derivative}:

\begin{equation}
\nabla_{\theta} L(\theta) = -2 \frac{1}{|B|} \left[ \sum_{a \in A^*} \log \pi_{\theta_{\textsubscript{Augmented}}} - \log \pi_{\theta_{\textsubscript{Agent}}} \right] \sum_{a \in A^*} \nabla_{\theta} \log \pi_{\theta_{\textsubscript{Agent}}}
\label{eq:main-text-squared-loss-derivative}
\end{equation}

\textbf{Equivalency of the Squared Difference Loss to Policy Gradient Optimization.} Minimizing the loss function described in Equation \ref{eq:main-text-dap} is equivalent to maximizing the expected reward. To show this equivalency, we follow Fialková et al.~\cite{fialkova_libinvent_2022} and start with the following objective, where $R$ is the reward function (Equation \ref{eq:main-text-expected-reward}):

\begin{equation}
J(\theta) = \mathbb{E}_{a_t \sim \pi_\theta} \left[ \sum_{t=0}^{T} R(a_t, s_t) \right]
\label{eq:main-text-expected-reward}
\end{equation}

Following the REINFORCE~\cite{williams_simple_1992} algorithm and applying the log-derivative trick yields Equation \ref{eq:main-text-log-derivative-trick} for the gradient:

\begin{equation}
\nabla_{\theta} J(\theta) = \mathbb{E}_{a_t \sim \pi_{\theta}} \left[ \sum_{t=0}^{T} R(a_t, s_t) \nabla_{\theta} \log \pi_{\theta}(a_t | s_t) \right]
\label{eq:main-text-log-derivative-trick}
\end{equation}

Computing the expectation is intractable and is instead approximated using the mean of a sampled batch, $B$, of SMILES constructed by choosing actions, $ a \in A^* $. Further noting that $ \log \pi_{\theta}(a_t | s_t) = \log \pi_{\theta_{\textsubscript{Agent}}} $ yields Equation \ref{eq:main-text-equivalency}:

\begin{equation}
\nabla_{\theta} J(\theta) = \frac{1}{|B|} \left[ \sum_{t=0}^{T} \sum_{a \in A^*} R(a_t, s_t) \nabla_{\theta} \log \pi_{\theta_{\textsubscript{Agent}}} \right]
\label{eq:main-text-equivalency}
\end{equation}

Finally, the reward is defined as $R(a_t, s_t) = \log \pi_{\theta_{\textsubscript{Augmented}}} - \log \pi_{\theta_{\textsubscript{Agent}}} $. The corresponding gradient expression (Equation \ref{eq:main-text-equivalency}) is now equivalent to the gradient of the loss function (Equation \ref{eq:main-text-squared-loss-derivative}) up to a constant factor. Further details on the derivation and algorithm is in Appendix \ref{loss-proof}.

\textbf{SMILES Augmentation.} SMILES are non-injective and yield different sequence representations given a different atom numbering in the molecular graph, i.e., augmented SMILES. SMILES-based molecular generative models have taken advantage of this to train performant models under low-data regimes, e.g., by artificially increasing the dataset size via data augmentation~\cite{moret_generative_2020}, and to increase chemical space generalizability~\cite{arus-pous_randomized_2019} by training a Prior model on augmented SMILES. Similar to Bjerrum et al.~\cite{bjerrum_faster_2023}, we reuse scores obtained from the oracle to update the Agent multiple times by passing different augmented SMILES representations.

\textbf{Experience Replay.} Experience replay is implemented in REINVENT as a buffer that stores a pre-defined maximum number of the highest rewarding SMILES sampled so far (100 in this work). Usually, during each sampling, a subset of the buffer is replayed to the Agent~\cite{blaschke_reinvent_2020}. In our proposed method, all SMILES in the buffer are augmented and using their corresponding reward, the Agent is updated multiple times according to the loss function given in Equation \ref{eq:main-text-dap}.

\textbf{Selective Memory Purge.} Blaschke et al.~\cite{blaschke_memory-assisted_2020} introduced memory-assisted RL to enforce diverse sampling in REINVENT via diversity filters (DFs). During the generative process, the scaffolds of sampled molecules are stored in 'buckets' with pre-defined and limited size. Once a bucket has been fully populated, further sampling of the same scaffold results in zero reward. We incorporate this heuristic in our proposed method called Selective Memory Purge to enforce diversity. At every epoch, the replay buffer is purged of any scaffolds that are penalized by the DF. The effect is that each augmentation round only updates the Agent with scaffolds that still receive reward, preventing the Agent from becoming myopic and leading to sub-optimal convergence.

\section{Results \& Discussion}
We designed three experiments to assess our method. First, we explicitly demonstrate the importance of experience replay and identify optimal parameters for Augmented Memory using the Aripiprazole Similarity experiment. Next, we benchmark its performance on the Practical Molecular Optimization (PMO)~\cite{gao_sample_2022} benchmark containing 23 tasks. Lastly, we demonstrate the practical applicability of our method on a Dopamine Type 2 Receptor (DRD2) drug discovery case study.

\textbf{Baselines.} In experiments 1 and 3, the baseline algorithms include REINVENT~\cite{olivecrona_molecular_2017, blaschke_reinvent_2020}, Augmented Hill Climbing (AHC)~\cite{thomas_augmented_2022}, Best Agent Reminder (BAR)~\cite{atance2022novo}, and Double Loop RL\cite{bjerrum_faster_2023} as all algorithms are formulated using REINVENT's~\cite{olivecrona_molecular_2017, blaschke_reinvent_2020} loss function with shared hyperparameters. Thus, the experiments isolate the effect of SMILES augmentation and experience replay on sample efficiency. Moreover, SMILES-based models are strong baselines as demonstrated in the GuacaMol~\cite{brown_guacamol_2019}, MOSES~\cite{polykovskiy_molecular_2020}, and PMO~\cite{gao_sample_2022} benchmarks. In the PMO benchmarking experiment, we compare our method's performance to diverse models. The Appendix includes further details on the dataset, hyperparameters, and ablation studies.

\begin{figure}[t]
\centering
\includegraphics[width=\linewidth]{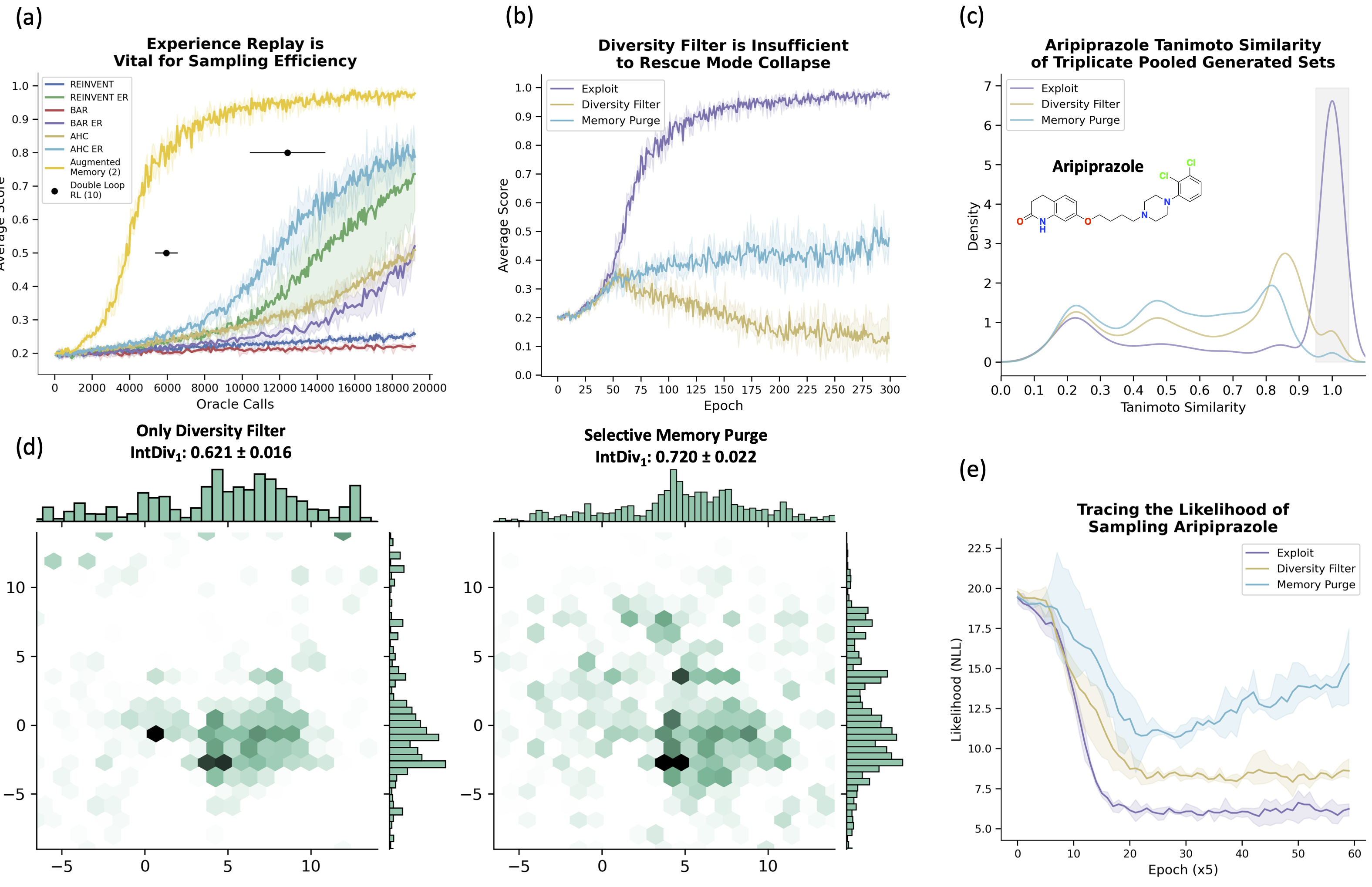} 
\caption{Augmented Memory and Selective Memory Purge significantly improve sample efficiency and enable diverse sampling. The shaded region represents the minimum and maximum scores across triplicate runs. (a) Comparing sample efficiency of on-policy algorithms. Experience Replay (ER) improves all base algorithms. The values for Double Loop RL~\cite{bjerrum_faster_2023} are taken from the original paper as the code is not released. The black dots are the mean at 0.5 and 0.8 and the standard deviation across triplicate runs. (b) The average score for aripiprazole similarity. In the Diversity Filter and Memory Purge experiments, scores of 0 are given if the Agent repeatedly samples the same scaffold. (c) Pooled Tanimoto similarities. Memory purge rediscovers aripiprazole and has a flatter distribution, suggesting increased exploration. (d) UMAP~\cite{mcinnes_umap_2020} and IntDiv1~\cite{polykovskiy_molecular_2020} metric showing qualitatively and quantitatively increased exploration using Memory Purge. The plots were generated using ChemCharts~\cite{chemcharts} (e) The negative log-likelihood of sampling aripiprazole across the full generative experiments.}
\label{fig:aripiprazole}
\end{figure}

\subsection{Aripiprazole Similarity}
The aripiprazole similarity task is from the GuacaMol benchmark~\cite{brown_guacamol_2019} and the objective is to successfully sample aripiprazole. This experiment was used to demonstrate the importance of experience replay and compare Augmented Memory to existing policy-based algorithms. As the code for Double Loop RL is not released, we took the values reported in the their paper which holds as the method was also built directly on REINVENT~\cite{olivecrona_molecular_2017, blaschke_reinvent_2020}, uses the same pre-trained Prior, and hyperparameters. Moreover, in the studies presenting AHC~\cite{thomas_augmented_2022} and BAR~\cite{atance2022novo}, experience replay was not used but we provide an implementation and further compare their performance.

\textbf{Experience Replay is Vital for Sample Efficiency.} We first identified the optimal number of augmentation rounds for Augmented Memory as two for training stability. Increasing the augmentation rounds can further improve sample efficiency but can lead to mode collapse (Appendix \ref{appendix:tolerability}). Next, we compare REINVENT~\cite{olivecrona_molecular_2017, blaschke_reinvent_2020}, AHC~\cite{thomas_augmented_2022}, BAR~\cite{atance2022novo}, and Double Loop RL~\cite{bjerrum_faster_2023} with our method. Figure \ref{fig:aripiprazole} plots the number of oracle calls to explicitly highlight the computational budget. Augmented Memory significantly outperforms all other algorithms and reaches a score of 0.8 with 6,144 oracle calls (average over 100 replicates). Double Loop RL~\cite{bjerrum_faster_2023} uses experience replay and is the second most sample efficient algorithm and reaches a score of 0.8 after 12,416 $\pm$ 1,984 oracle calls (as stated in their paper), which is twice the number of oracle calls required compared to our method. Moreover, the key observation we convey is that experience replay improves upon the base algorithm in all cases (Figure \ref{fig:aripiprazole}). For example, AHC~\cite{thomas_augmented_2022} with the newly implemented experience replay reaches a score of 0.8, but with more than triple the oracle calls (19,200). Our observations around experience replay are supported by previous works~\cite{korshunova_generative_2022, blaschke_reinvent_2020}. Finally, we show that augmentation is crucial for the enhanced sample efficiency in Appendix \ref{appendix:ablation}.

\textbf{Selective Memory Purge Enables Diverse Sampling while Retaining Efficiency.} Figure \ref{fig:aripiprazole} demonstrates the enhanced sample efficiency of Augmented Memory but real-world applications of molecular generative models require the ability to sample diverse solutions. While aripiprazole is inherently an exploitation task, it can be framed as an exploration task if the goal is rephrased as: rediscover the target molecule and generate similar molecules. Using this formulation, we design experiments to prove that Augmented Memory can achieve diverse sampling. Figure \ref{fig:aripiprazole} shows the training plot across three methods: pure exploitation where diversity is not enforced, exploration using a diversity filter (DF)~\cite{blaschke_memory-assisted_2020}, and Selective Memory Purge. In the pure exploitation scenario, aripiprazole is rediscovered quickly (score of 1.0). In the DF experiment where a score of 0 is assigned for scaffolds sampled more than 25 times, mode collapse is observed (Figure \ref{fig:aripiprazole}). By contrast, Selective Memory Purge maintains a moderate average score. The results from triplicate experiments were pooled to investigate the density of aripiprazole similarities (Figure \ref{fig:aripiprazole}). As expected, in the pure exploitation scenario, most molecules are aripiprazole (Tanimoto similarity of 1.0). DF and Selective Memory Purge both enforce a wider distribution of similarities, but to varying degrees. In the shaded region (rediscovery score), Selective Memory Purge only shows a small density relative to DF. Moreover, Selective Memory Purge shows a flatter distribution of similarities. These observations demonstrate that Selective Memory Purge rediscovers the target molecule and enforces increased exploration compared to DF. To investigate this further, the same pooled dataset was embedded using Uniform Manifold Approximation and Projection (UMAP)~\cite{mcinnes_umap_2020} to visualize the chemical space. Qualitatively and quantitatively, Selective Memory Purge covers a larger chemical space (Figure \ref{fig:aripiprazole}). The internal diversity (IntDiv1) metric was calculated as proposed in the MOSES benchmark~\cite{polykovskiy_molecular_2020}, and measures the diversity within a set of generated molecules. Finally, we save the Agent states at every 5 epochs across the entire generative run and trace the NLL of sampling aripiprazole (Figure \ref{fig:aripiprazole}). It is evident that Selective Memory Purge can discourage sampling of the target molecule more effectively than only using a DF. Importantly, the NLL also diverges, suggesting that the Agent is increasingly moving to chemical space dissimilar to aripiprazole as the generative experiment progresses.

\subsection{Practical Molecular Optimization (PMO) Benchmark}

\begin{table}[ht!]
\scriptsize
\centering
\caption{Performance of Augmented Memory, REINVENT~\cite{olivecrona_molecular_2017, blaschke_reinvent_2020}, AHC~\cite{thomas_augmented_2022}, and BAR~\cite{atance2022novo} on the PMO benchmark~\cite{gao_sample_2022}. The mean and standard deviation of the AUC Top-10 is reported. The values obtained for REINVENT differ slightly from the PMO paper as we performed 10 independent runs compared to 5. Best performance is bolded and is relative to all models in the benchmark. * denotes superior performance to REINVENT but not overall, compared to other models in the benchmark. We note however, that we take the AUC Top-10 values for the other models as is from the PMO paper. If they were re-run with 10 different seeds (instead of 5), the values may decrease as was observed for REINVENT.}
\begin{tabular}{|c|c|c|c|c|c|c|}
\toprule
\noalign{\vskip 2pt}
\textbf{Benchmark} & \textbf{Augmented} & \textbf{REINVENT} & \textbf{AHC} & \textbf{BAR} & \textbf{AHC} & \textbf{BAR} \\[1pt]
\textbf{Task} & \textbf{Memory} & & \textbf{Replay} & \textbf{Replay} & & \\[1pt]
\hline
\noalign{\vskip 2pt}
albuterol\_similarity & \textbf{0.913 $\pm$ 0.009} & 0.871 $\pm$ 0.031 & 0.792 $\pm$ 0.030 & 0.700 $\pm$ 0.024 & 0.745 $\pm$ 0.024 & 0.633 $\pm$ 0.031\\[1pt]
amlodipine\_mpo & \textbf{0.691 $\pm$ 0.047} & 0.657 $\pm$ 0.025 & 0.596 $\pm$ 0.023 & 0.538 $\pm$ 0.019 & 0.578 $\pm$ 0.012 & 0.523 $\pm$ 0.006\\[1pt]
celecoxib\_rediscovery & \textbf{0.796 $\pm$ 0.008} & 0.717 $\pm$ 0.048 & 0.697 $\pm$ 0.029 & 0.563 $\pm$ 0.043 & 0.583 $\pm$ 0.070 & 0.437 $\pm$ 0.025\\[1pt]
deco\_hop & 0.658 $\pm$ 0.024 & 0.672 $\pm$ 0.052 & 0.650 $\pm$ 0.030 & 0.589 $\pm$ 0.010 & 0.632 $\pm$ 0.032 & 0.579 $\pm$ 0.008\\[1pt]
drd2 & \textbf{0.963 $\pm$ 0.006}$^{*}$ & 0.939 $\pm$ 0.012 & 0.913 $\pm$ 0.011 & 0.916 $\pm$ 0.012 & 0.912 $\pm$ 0.009 & 0.899 $\pm$ 0.027\\[1pt]
fexofenadine\_mpo & \textbf{0.859 $\pm$ 0.009} & 0.783 $\pm$ 0.021 & 0.747 $\pm$ 0.004 & 0.708 $\pm$ 0.010 & 0.749 $\pm$ 0.005 & 0.692 $\pm$ 0.009\\[1pt]
gsk3b & \textbf{0.881 $\pm$ 0.021} & 0.870 $\pm$ 0.026 & 0.819 $\pm$ 0.025 & 0.744 $\pm$ 0.021 & 0.800 $\pm$ 0.021 & 0.686 $\pm$ 0.068\\[1pt]
isomers\_c7h8n2o2 & 0.853 $\pm$ 0.087 & 0.856 $\pm$ 0.042 & 0.682 $\pm$ 0.037 & 0.741 $\pm$ 0.064 & 0.631 $\pm$ 0.084 & 0.713 $\pm$ 0.058\\[1pt]
isomers\_c9h10n2o2pf2cl & \textbf{0.736 $\pm$ 0.051}$^{*}$ & 0.641 $\pm$ 0.038 & 0.276 $\pm$ 0.133 & 0.612 $\pm$ 0.054 & 0.191 $\pm$ 0.096 & 0.508 $\pm$ 0.066\\[1pt]
jnk3 & \textbf{0.739 $\pm$ 0.110} & 0.723 $\pm$ 0.147 & 0.649 $\pm$ 0.056 & 0.555 $\pm$ 0.089 & 0.616 $\pm$ 0.092 & 0.511 $\pm$ 0.092\\[1pt]
median1 & 0.326 $\pm$ 0.013 & 0.368 $\pm$ 0.011 & 0.346 $\pm$ 0.008 & 0.286 $\pm$ 0.007 & 0.338 $\pm$ 0.014 & 0.269 $\pm$ 0.011\\[1pt]
median2 & \textbf{0.291 $\pm$ 0.008}$^{*}$ & 0.279 $\pm$ 0.005 & 0.273 $\pm$ 0.005 & 0.218 $\pm$ 0.008 & 0.265 $\pm$ 0.005 & 0.199 $\pm$ 0.006\\[1pt]
mestranol\_similarity & \textbf{0.750 $\pm$ 0.049} & 0.637 $\pm$ 0.041 & 0.599 $\pm$ 0.031 & 0.463 $\pm$ 0.027 & 0.561 $\pm$ 0.022 & 0.444 $\pm$ 0.017\\[1pt]
osimertinib\_mpo & \textbf{0.855 $\pm$ 0.004} & 0.836 $\pm$ 0.007 & 0.810 $\pm$ 0.003 & 0.789 $\pm$ 0.012 & 0.809 $\pm$ 0.002 & 0.792 $\pm$ 0.004\\[1pt]
perindopril\_mpo & \textbf{0.613 $\pm$ 0.015} & 0.561 $\pm$ 0.019 & 0.487 $\pm$ 0.012 & 0.468 $\pm$ 0.012 & 0.482 $\pm$ 0.008 & 0.455 $\pm$ 0.011\\[1pt]
qed & \textbf{0.942 $\pm$ 0.000} & 0.941 $\pm$ 0.000 & 0.941 $\pm$ 0.000 & 0.939 $\pm$ 0.003 & 0.941 $\pm$ 0.000 & 0.932 $\pm$ 0.007\\[1pt]
ranolazine\_mpo\_mpo & \textbf{0.801 $\pm$ 0.006} & 0.768 $\pm$ 0.008 & 0.721 $\pm$ 0.00 & 0.704 $\pm$ 0.017 & 0.722 $\pm$ 0.008 & 0.700 $\pm$ 0.021\\[1pt]
scaffold\_hop & \textbf{0.567 $\pm$ 0.008} & 0.556 $\pm$ 0.019 & 0.535 $\pm$ 0.007 & 0.477 $\pm$ 0.010 & 0.525 $\pm$ 0.008 & 0.464 $\pm$ 0.005\\[1pt]
sitagliptin\_mpo & \textbf{0.284 $\pm$ 0.050}$^{*}$ & 0.049 $\pm$ 0.067 & 0.022 $\pm$ 0.008 & 0.126 $\pm$ 0.049 & 0.028 $\pm$ 0.011 & 0.070 $\pm$ 0.020\\[1pt]
thiothixene\_rediscovery & \textbf{0.550 $\pm$ 0.041}$^{*}$ & 0.531 $\pm$ 0.036 & 0.519 $\pm$ 0.012 & 0.396 $\pm$ 0.011 & 0.467 $\pm$ 0.032 & 0.347 $\pm$ 0.013\\[1pt]
troglitazone\_rediscovery & \textbf{0.540 $\pm$ 0.048} & 0.428 $\pm$ 0.028 & 0.409 $\pm$ 0.020 & 0.301 $\pm$ 0.007 & 0.371 $\pm$ 0.019 & 0.279 $\pm$ 0.007\\[1pt]
valsartan\_smarts & 0.000 $\pm$ 0.000 & 0.091 $\pm$ 0.273 & 0.000 $\pm$ 0.000 & 0.000 $\pm$ 0.000 & 0.000 $\pm$ 0.000 & 0.000 $\pm$ 0.000\\[1pt]
zaleplon\_mpo & \textbf{0.394 $\pm$ 0.026} & 0.269 $\pm$ 0.083 & 0.072 $\pm$ 0.032 & 0.319 $\pm$ 0.033 & 0.047 $\pm$ 0.013 & 0.294 $\pm$ 0.014\\[1pt]
\midrule
\noalign{\vskip 2pt}
Sum of AUC Top-10 (↑) & \textbf{15.002} & 14.016 & 12.555 & 12.152 & 11.993 & 11.426\\
PMO Rank ($n/29$) & \textbf{1} & 2 & 7 & 9 & 11 & 14\\[1pt]
\noalign{\vskip 2pt}
\bottomrule
\end{tabular}
\label{table:pmo}
\end{table}

The main motivation of our method is to improve sample efficiency. This would enable molecular generative models to explicitly optimize more expensive oracles which can afford increased predictive accuracy. We benchmark our method on the PMO benchmark proposed by Gao et al.~\cite{gao_sample_2022} which restricts the number of oracle calls to 10,000 and encompasses 23 tasks. The metric used is the Area Under the Curve (AUC) for the top 10 molecules. We note that Thomas et al.~\cite{thomas2022re} proposed a modified AUC Top-10 metric that incorporates diversity, but we omit comparison as the formulation can be subjective. The current AUC Top-10 metric assesses sample efficiency which is our focus. In the original PMO paper, REINVENT\cite{olivecrona_molecular_2017} (with experience replay) is the most sample efficient model. We compare our method directly to REINVENT, BAR~\cite{atance2022novo}, and AHC~\cite{thomas_augmented_2022} which reports improved sample efficiency compared to REINVENT and is open-sourced. We also add experience replay to BAR and AHC to further highlight its importance for sample efficiency. For a more statistically convincing comparison, we perform 10 independent runs (using 10 different seeds) compared to 5 used in the original PMO paper as the authors benchmarked 25 models, which imposed a significant computational cost. The optimal hyperparameters for REINVENT and AHC were used as provided in the PMO repository. We perform hyperparameter optimization for BAR following the PMO protocol (Appendix \ref{appendix:pmo-tuning}) and Augmented Memory was run using REINVENT's optimal hyperparameters. The results show Augmented Memory significantly outperforms all methods and achieves superior performance to REINVENT across 19/23 benchmark tasks (Table \ref{table:pmo}). Moreover, the results reinforce the importance of experience replay as it improves the sample efficiency of both BAR and AHC, although neither outperform REINVENT. Finally, in the PMO paper~\cite{gao_sample_2022}, models were ranked based on the sum of the total AUC Top-10 and adjacently ranked models typically differ by 0.3-0.5. Augmented Memory outperforms REINVENT by 0.986 AUC Top-10 and yields a new state-of-the-art performance on the PMO benchmark. 

\begin{figure}
\centering
\includegraphics[width=1.0\columnwidth]{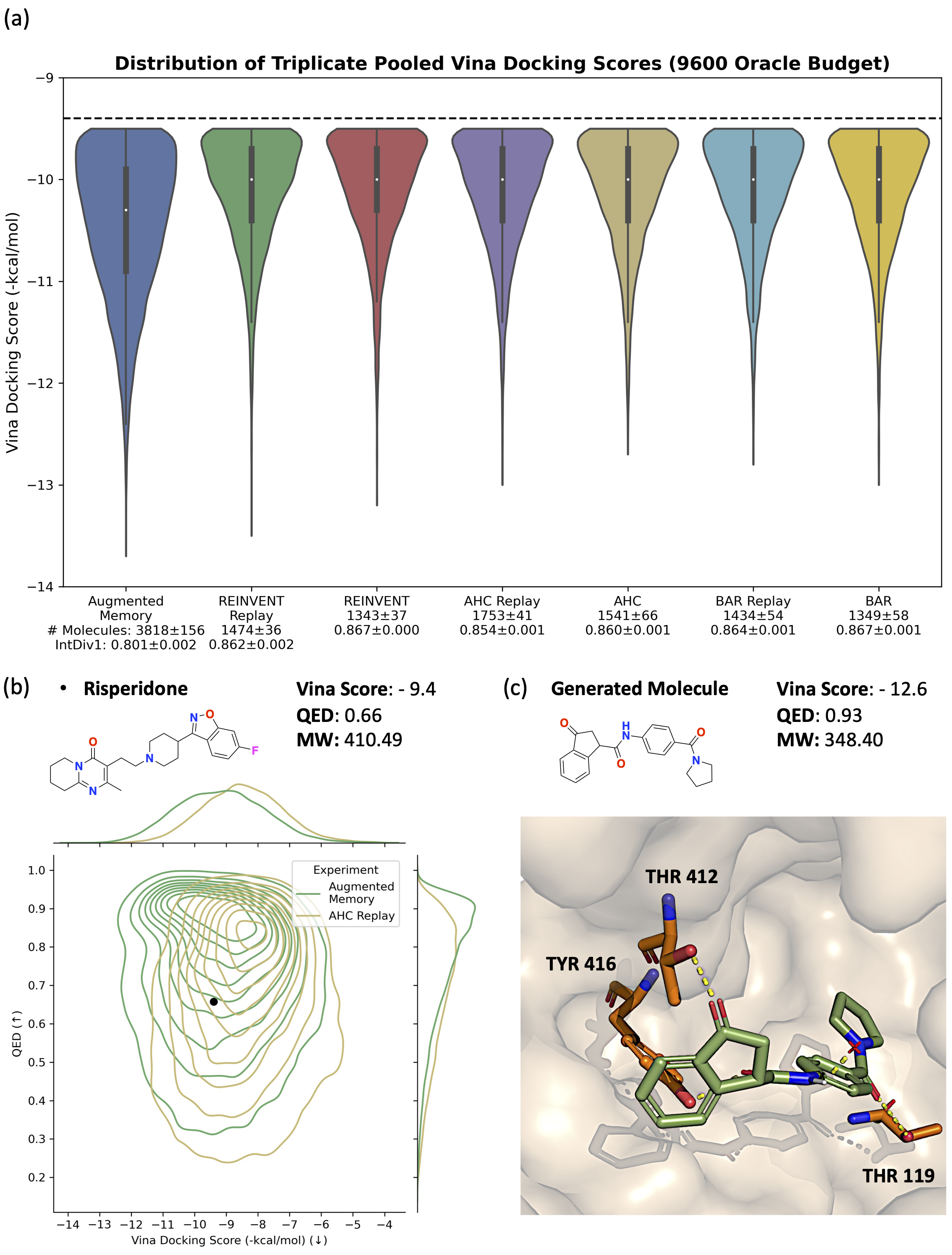} 
\caption{Dopamine type 2 receptor (DRD2) molecular docking case study. PDB ID: 6CM4. (a) Docking scores distribution of all compared algorithms. (b) Augmented Memory jointly optimizes QED and Vina docking score, demonstrating the ability to perform MPO. (c) Binding pose of a generated molecule using Augmented Memory. The three components in the objective function: MW < 500, QED, and Vina docking score are all optimized.}
\label{fig:drd2}
\end{figure}

\subsection{Dopamine Type 2 Receptor (DRD2) Case Study}
To prove that Augmented Memory can perform MPO, we formulate a case study to generate potential dopamine type 2 receptor (DRD2) inhibitors~\cite{wang2018structure} by explicitly optimizing molecular docking scores (Figure \ref{fig:drd2}). For accessibility and reproducibility, we use the open-source AutoDock Vina~\cite{trott2010autodock} for docking. A well-known failure mode of docking algorithms is they reward lipophilic molecules, e.g., possessing many carbon atoms, which can be promiscuous binders~\cite{arnott2012influence, nigam2022parallel}. Bjerrum et al.~\cite{bjerrum_faster_2023} consider this and enforced molecules to possess a molecular weight (MW) < 500 Da but this is insufficient in preventing exploitation of the docking algorithm as we show in Appendix \ref{appendix:exploit-vina}. Following Guo et al.~\cite{guo_dockstream_2021}, we design the MPO as follows: MW < 500 Da, maximize QED~\cite{bickerton_quantifying_2012}, and minimize the Vina docking score, for chemical plausibility. AutoDock Vina is a relatively expensive oracle and we impose a computational budget of 9,600 oracle calls, similar to the 10,000 oracle calls enforced in the PMO~\cite{gao_sample_2022} benchmark. We compare Augmented Memory, REINVENT~\cite{olivecrona_molecular_2017, blaschke_reinvent_2020}, AHC~\cite{thomas_augmented_2022}, and BAR~\cite{atance2022novo} as the optimization algorithms. To mimic a real-world drug discovery pipeline that discards unpromising molecules, we pool the results from triplicate experiments with the following filter: MW < 500 Da, QED > 0.4 (the DRD2 drug molecule, risperidone, has a QED of 0.66), and Vina docking score < -9.4 (risperidone's score). Figure \ref{fig:drd2} shows the docking scores distribution with the number of molecules passing the filter and the IntDiv1~\cite{polykovskiy_molecular_2020} score annotated. Firstly, experience replay improves all base algorithms, further reinforcing its importance. Secondly, all algorithms with the exception of Augmented Memory perform similarly. Compared to AHC with experience replay, which is the second most sample efficient algorithm, Augmented Memory generates over 2,000 more molecules with a better docking score than risperidone, with a small trade-off in diversity (IntDiv1 of 0.801). We emphasize that AHC with experience replay does not even generate 2,000 molecules passing the filter. To further prove the optimization capability, Figure \ref{fig:drd2} shows a contour plot of the QED-Vina score distribution for Augmented Memory and AHC with experience replay. It is clear that the joint QED-Vina score distribution for Augmented Memory is shifted to higher QED values and lower Vina scores. The black dot is risperidone and the bulk density of AHC does not possess a better docking score. Finally, Figure \ref{fig:drd2} shows an example binding pose of a molecule generated using Augmented Memory. We highlight that the chemical plausibility of the structure is enforced precisely because MW and QED are also included in the MPO objective, thus representing a more realistic case study.

\section{Conclusion}
In this work, we explicitly show that experience replay is vital for sample efficiency. We propose Augmented Memory which capitalizes on this observation and applies SMILES augmentation on the replay buffer to update the Agent multiple times per oracle call. Compared to existing algorithms, Augmented Memory significantly improves sample efficiency and is able to sample diverse solutions using the newly proposed Selective Memory Purge heuristic. We benchmark Augmented Memory on the PMO benchmark~\cite{gao_sample_2022} and achieve a new state-of-the-art performance, outperforming the previous state-of-the-art on 19/23 tasks and by a total sum of 0.986 AUC Top-10. Next, we show the practical application of Augmented Memory by mimicking a more realistic drug discovery task. Our method significantly outperforms existing algorithms, as assessed by the property profile of the generated molecules, and can perform MPO. We note that in particularly sparse reward landscapes~\cite{korshunova_generative_2022}, the enhanced sample efficiency of Augmented Memory may be diminished as it becomes more difficult to populate the replay buffer with high rewarding molecules. Future work will investigate this scenario thoroughly and algorithmic modifications to couple additional local chemical space exploration around high rewarding molecules may better handle sparsity. Finally, this work opens up future integration of Augmented Memory with curriculum learning~\cite{guo_improving_2022}, the use of more expensive oracles given a limited computational budget, and further provides insights into experience replay for molecular generative models.

\section*{Acknowledgement}

This publication was created as part of NCCR Catalysis (grant number 180544), a National Centre of Competence in Research funded by the Swiss National Science Foundation.

\citestyle{nature}
\bibliographystyle{unsrtnat}  
\bibliography{bibliography}

\appendix
\section{Tolerability to Augmentation Rounds}
\label{appendix:tolerability}

\begin{figure}[ht]
\centering
\includegraphics[width=1.0\columnwidth]{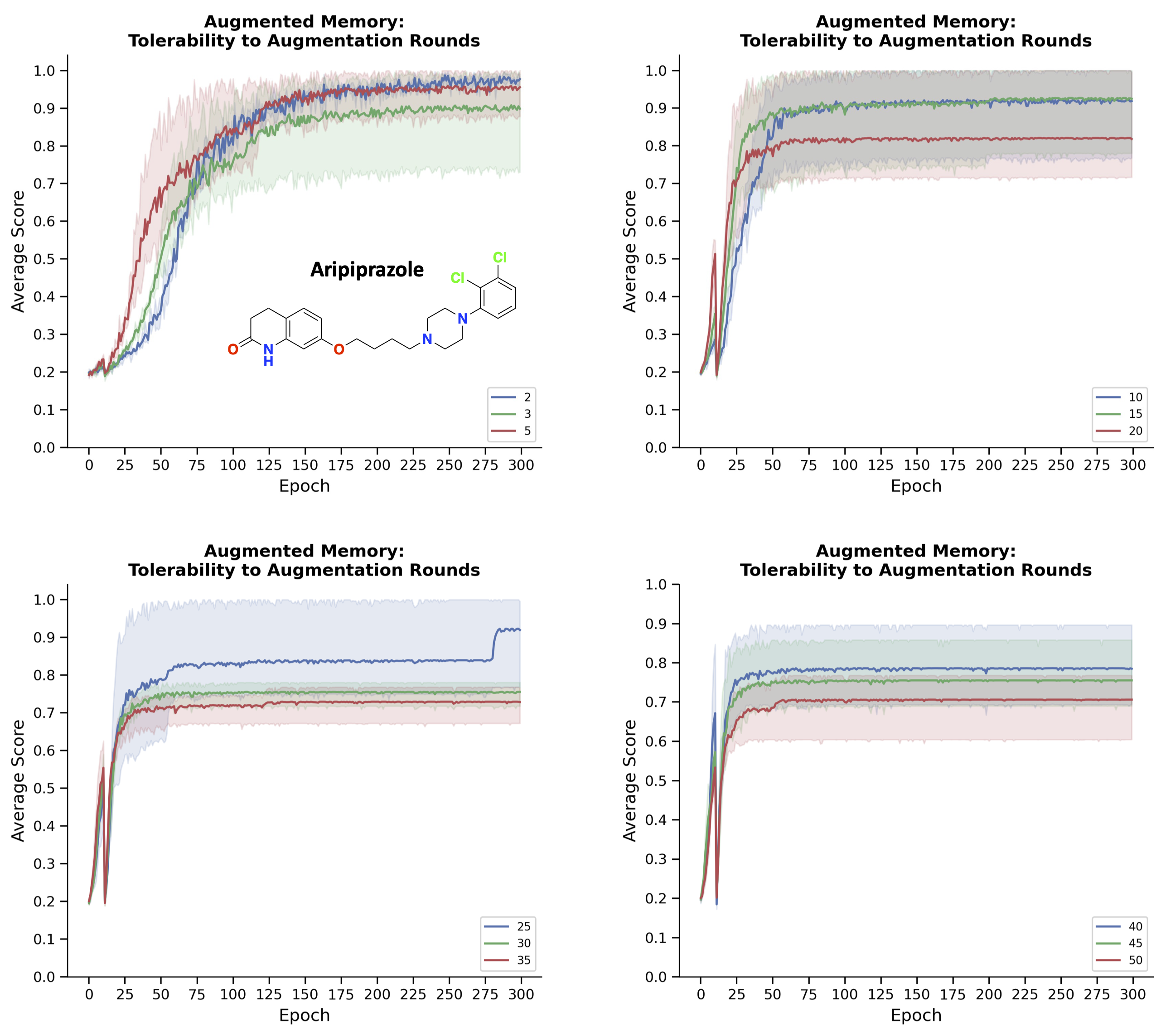} 
\caption{Identifying the optimal augmentation rounds using aripiprazole similarity. The shaded region represents the minimum and maximum scores across triplicate runs.}
\label{fig:appendix-augmentation-rounds}
\end{figure}

Similar to Esben et al.~\cite{bjerrum_faster_2023} in their proposed Double Loop RL algorithm, increasing the number of augmentation rounds increases susceptibility to mode collapse (Figure \ref{fig:appendix-augmentation-rounds}). We used the aripiprazole similarity task to perform a grid optimization and found two rounds to be optimal for stability. At three rounds, mode collapse is already observed with triplicate runs.

\section{Pure Exploitation: Robustness of 2 Augmentation Rounds}

\begin{table}[ht]
    \centering
    \caption{Robustness experiments: stability of two augmentation rounds. 100 replicates of aripiprazole similarity was performed using 2 augmentation rounds and the epoch number to reach various average scores are presented. The values for Double Loop RL~\cite{bjerrum_faster_2023} are for 10 augmentations which the authors state to be most stable}
    \begin{tabular}{|l|c|c|c|c|}
        \hline
        Average Score & Mean & Minimum & Maximum & Double Loop RL \\[2pt]
        & Epochs & Epochs & Epochs & Epochs \\[2pt]
        \hline
        0.5 & $64 \pm 6$ & 51 & 79 & $93 \pm 9$ \\[2pt]
        \hline
        0.8 & $96 \pm 18$ & 77 & 195 & $194 \pm 31$ \\[2pt]
        \hline
        0.9 & $122 \pm 17$ & 97 & 215 & did not report \\[2pt]
        \hline
    \end{tabular}
    \label{table: robustness}
\end{table}

Initial screening experiments identified two augmentation rounds to be optimal for training stability. We envisioned in \textbf{pure exploitation} scenarios where \textbf{Selective Memory Purge} is not used, mode collapse may be possible. The rationale being that the Agent is reinforced on the same replay buffer molecules. In the case where the entire replay buffer contains very similar or identical molecules, mode collapse may occur. \textbf{This is not an issue when using Selective Memory Purge as entries in the replay buffer would be removed, thus preventing the entire buffer containing the same molecules.} In most practical applications of molecular generative models, Selective Memory Purge should be used to achieve both exploration and exploitation. However, for full transparency, we report the stability of our proposed method in a pure exploitation scenario. The following insights will be informative if prospective users only want to generate one optimal solution in their generative experiment or want to reproduce the aripiprazole similarity experiment. To preemptively prevent mode collapse, we introduce "mode collapse guard" that purges the replay buffer if 70 percent (empirically we find this to be sufficient) of the buffer contains the exact same reward. For statistical rigour, we perform 100 replicates of aripiprazole similarity and present the results in Table \ref{table: robustness}. We follow Bjerrum et al.~\cite{bjerrum_faster_2023} and present statistics on the epochs it takes to reach various average scores (average Tanimoto similarity of the batch of sampled molecules to aripiprazole) of 0.5, 0.8, and 0.9. The results support the stability of our method even in pure exploitation scenarios. The "mode collapse guard" was activated 14 times across 100 replicates and in all cases except 1, prevents mode collapse. The one exception failed to rediscover aripiprazole (mode collapse at a Tanimoto similarity of 0.78). In practical applications, the experiment can be monitored and restarted from a check-point state. Moreover, we comment on the maximum epochs it takes to reach an average score of 0.8 and 0.9 which are, in both cases, more than 5 standard deviations from the mean, thus extremely rare. We compare these results to the performance reported by Bjerrum et al.~\cite{bjerrum_faster_2023} in their Double Loop RL work which is the second most sample-efficient algorithm. Their reported values to reach an average score of 0.8 is $194 \pm 31$ using 10 augmentation rounds across triplicate runs. We first note that it is unclear if running their algorithm for 100 replicates would still be stable as it is not open-sourced. Secondly, our worst performance, taking 195 epochs to reach an average score of 0.8 is essentially identical to their mean epoch of 194. Cross-referencing the mean it takes our method, we highlight that Augmented Memory is much more sample-efficient, as we find in the main results.

\section{Buffer Size Experiments and Reinforcing with Only Experience Replay}

\begin{figure}[ht]
\centering
\includegraphics[width=1.0\columnwidth]{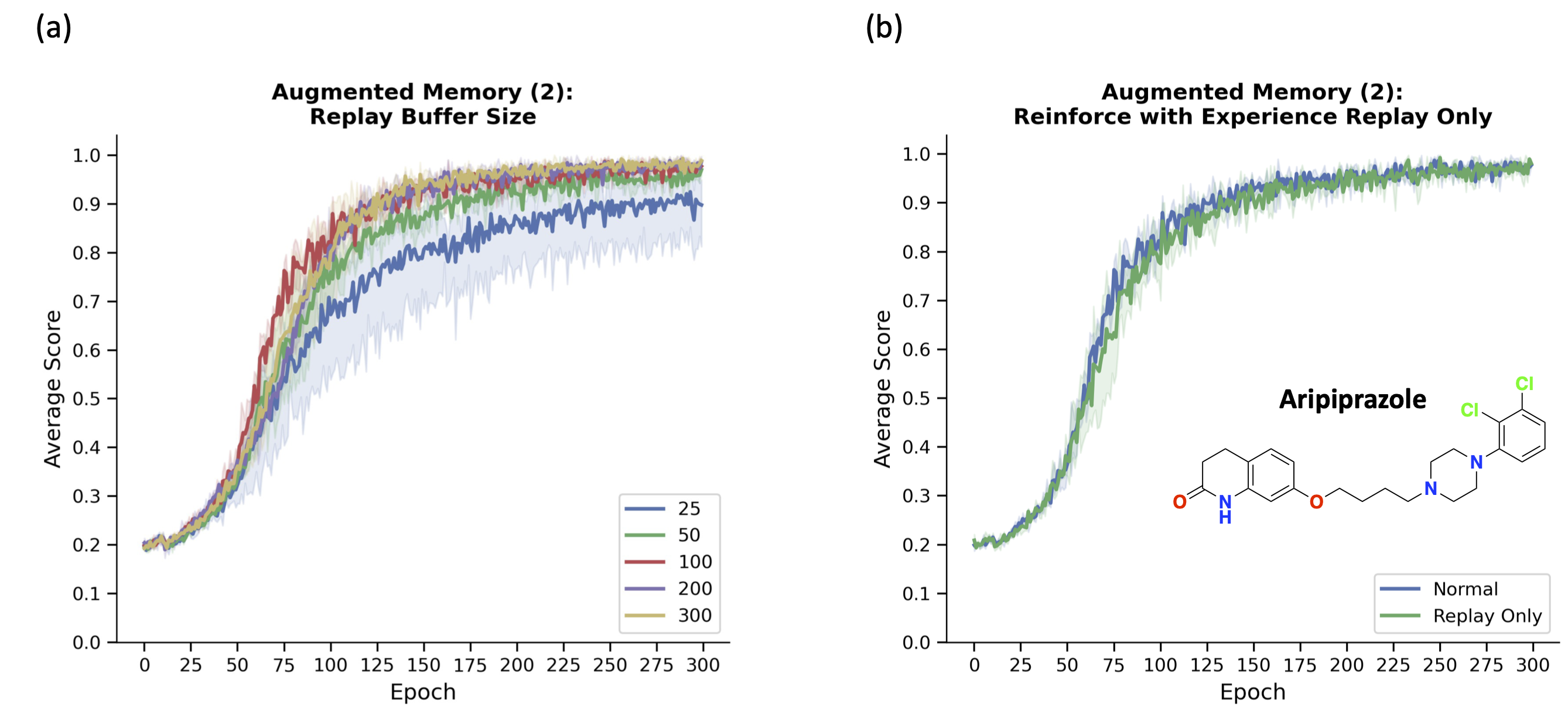} 
\caption{Investigating changes in the replay buffer size and reinforcing the Agent only with molecules stored in the replay buffer. The shaded region represents the minimum and maximum scores across triplicate runs.}
\label{fig:appendix-buffer}
\end{figure}

As Augmented Memory revolves around exploiting experience replay, we investigate the efficacy of our method when using different buffer sizes (Figure \ref{fig:appendix-buffer}). We again use the aripiprazole similarity task to assess the proposed changes. Interestingly, with the exception of a buffer size of 25, minimal difference is observed between buffer sizes. We posit that a buffer size of 25 is more susceptible to mode collapse as it is increasingly likely that the stored molecules are all identical or similar relative to having a larger buffer size. Conversely, our initial hypothesis was that a larger buffer size would decrease sample efficiency. The rationale is that relatively low rewarding molecules may be stored in the buffer and reinforcing on these low rewarding molecules could be counterproductive. Following experiments (Figure \ref{fig:appendix-buffer}), this was not the case, at least for the aripiprazole similarity task. Given that the differences in the buffer sizes result in minimal difference and that our hypothesis may be true for other objective functions, we decided to use a buffer size of 100 for main result experiments. Next, we were curious if reinforcing the Agent with only the molecules in the replay buffer would be possible. In these experiments, the sampled molecules in a given epoch were only used to reinforce the Agent once and no augmented forms were used to further reinforce the Agent. Interestingly, minimal difference is observed again (Figure \ref{fig:appendix-buffer}). Since the performance is similar, we hypothesize that using augmented forms of the sampled molecules would act to mitigate against mode collapse. This is in agreement with insights from Arús-Pous et al.~\cite{arus-pous_randomized_2019} that posit SMILES augmentation acts as a regularizer. Therefore, all main result experiments were performed using augmented SMILES from the sampled batch and the buffer.

\section{Ablation Study: SMILES Augmentation is a Regularizer}
\label{appendix:ablation}

\begin{table}[ht]
    \centering
    \caption{Stability without SMILES augmentation. 100 replicates of aripiprazole similarity was performed using 2 augmentation rounds (but without SMILES augmentation) and the epoch number to reach various average scores are presented. Failed runs did not reach the average score threshold. The epoch numbers for the runs with augmentation are shown in parenthesis for comparison.}
    \begin{tabular}{|l|c|c|c|c|}
        \hline
        Average Score & Mean & Minimum & Maximum & Failed Runs \\[2pt]
        & Epochs & Epochs & Epochs & \\[2pt]
        \hline
        0.5 & $69 (64) \pm 7 (6)$ & 53 (51) & 84 (79) & 0 \\[2pt]
        \hline
        0.8 & $115 (96) \pm 35 (18)$ & 74 (77) & 261 (194) & 3 \\[2pt]
        \hline
        0.9 & $154 (122) \pm 43 (17)$ & 94 (97) & 297 (215) & 9 \\[2pt]
        \hline
    \end{tabular}
    \label{table: no-smiles-augmentation}
\end{table}

The results in the main text show that experience replay is vital for sampling efficiency. In this section, the question we answer is: "can we just perform multiple rounds of Agent update with the entire replay buffer without SMILES augmentation?" If yes, then the benefits of Augmented Memory can be attributed to simply experience replay. The experimental design is as follows: using the aripiprazole similarity experiment, perform two rounds of Agent update using the entire buffer (size of 100) without SMILES augmentation. This mirrors the optimal parameters of Augmentation Memory of two augmentation rounds and a buffer size of 100. For statistical rigour, we perform 100 replicates and present the results in Table \ref{table: no-smiles-augmentation}. Compared to Augmented Memory, the average epochs it takes to reach an average score of 0.5, 0.8, and 0.9 is higher (values with augmentation are shown in parentheses and is from Table \ref{table: robustness}). Importantly, the standard deviation is also much higher, suggesting instability in the runs. This is further supported by some runs not reaching the 0.8 and 0.9 average score thresholds. While monitoring the sampling, we notice that the Agent repeatedly samples the same SMILES, indicating mode collapse. From a probabilistic perspective, the Agent negative log-likelihoods (NLLs) become focused on the replay buffer sequences, suggesting token-level memorization. Moreover, the minimum epochs for the 0.8 and 0.9 average score thresholds are lower than Augmented Memory. While seemingly suggesting higher efficacy compared to Augmented Memory, this observation instead further supports token-level memorization: if by chance the Agent finds favorable SMILES, it will focus only on those SMILES sequences. These insights are supported by previous work from Arús-Pous et al.~\cite{arus-pous_randomized_2019} which explored the effect of SMILES augmentation on the Prior's NLL on the training data. Specifically, they found that training a Prior without SMILES augmentation can cause token-level memorization, such that the NLL for the specific SMILES sequences in the training data are low. This decreases the generalizability of the trained Prior. Bjerrum et al.~\cite{bjerrum_faster_2023} in their Double Loop RL work also posit that reinforcing the Agent on augmented SMILES prevents sequence-wise mode collapse. Our results are in agreement and we show that SMILES augmentation is necessary to ensure the efficacy of Augmented Memory and is itself a regularizer. 

\section{Dopamine Type 2 Receptor (DRD2) Case Study: Exploiting AutoDock Vina}
\label{appendix:exploit-vina}

\begin{figure}[ht]
\centering
\includegraphics[width=1.0\columnwidth]{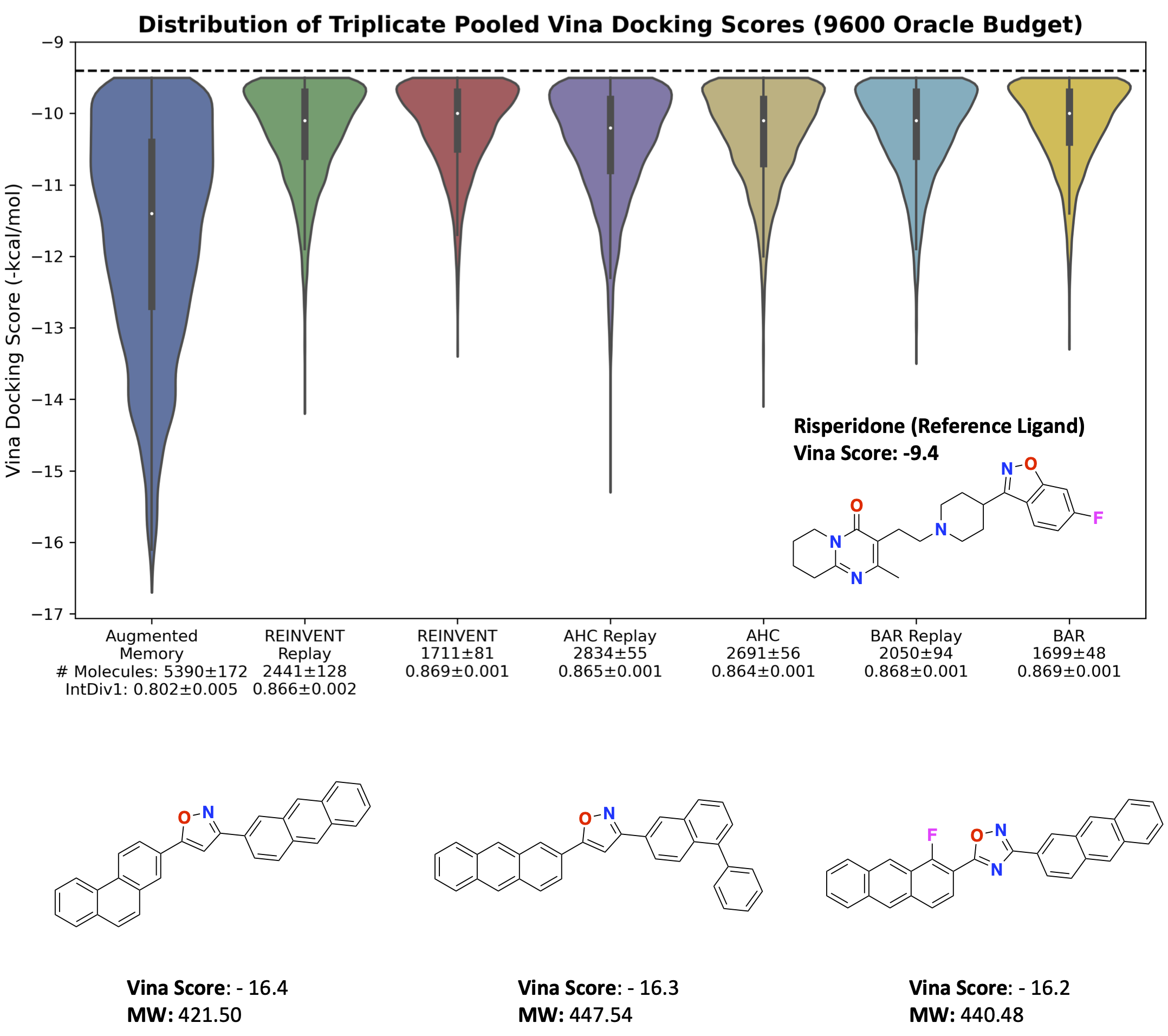} 
\caption{Dopamine type 2 receptor (DRD2) case study using the objective function: molecular weight < 500 Da and minimize Vina docking score. Augmented Memory significantly outperforms other algorithms. The generated molecules, however, are not realistic and shows that Augmented Memory can exploit objective functions in a sample-efficient manner.}
\label{fig:appendix-exploit-vina}
\end{figure}

This section elaborates on the statement that the experimental design of Bjerrum et al.~\cite{bjerrum_faster_2023} in their Double Loop RL work is insufficient in preventing AutoDock Vina~\cite{trott2010autodock} exploitation. Specifically, the drug discovery case study to design potential dopamine type 2 receptor (DRD2) inhibitors was performed using the following objective function: molecular weight (MW) < 500 Da, maximize QED, and minimize docking score. This is in contrast to the objective function proposed by Bjerrum et al.:  molecular MW < 500 Da and minimize docking score. We perform a set of experiments comparing the sample efficiency of alternative algorithms including REINVENT~\cite{olivecrona_molecular_2017, blaschke_reinvent_2020}, Best Agent Reminder (BAR)~\cite{atance2022novo}, and Augmented Hill Climbing (AHC)~\cite{thomas_augmented_2022} using this simplified objective function. Similar to the main result experiments, we ran all experiments enforcing an oracle budget of 9,600 calls and show the distribution of Vina scores from triplicate pooled runs. The following filter was applied: MW < 500 Da and Vina score < -9.4 (the Vina score of the reference drug molecules, risperidone). It is evident that Augmented Memory significantly outperforms other algorithms, generating drastically better docking scores. Moreover, there is minimal difference between the performances of the other algorithms. However, we investigate the property profile of the generated molecules in the Augmented Memory experiments and show that the Agent exploits AutoDock Vina in rewarding lipophilic molecules, i.e., all top scoring molecules have extensive aromatic carbon rings (Figure \ref{fig:appendix-exploit-vina}). These molecules, while possessing excellent Vina scores, are not realistic. As we emphasize the usability of Augmented Memory on more realistic case studies to encourage practical applications, we show the set of experiments which also enforce QED~\cite{bickerton_quantifying_2012} in the main results. QED ensures generated molecules are "drug-like". We end this section by emphasizing that the ability of Augmented Memory to exploit AutoDock Vina is not a weakness and rather, further proves its ability for sample-efficient optimization.

\section{Aripiprazole and DRD2 Prior and Hyperparameters}
\label{appendix-hyperparameters}

The \emph{random.prior.new} pre-trained Prior was used from the REINVENT 2.0~\cite{blaschke_reinvent_2020} repository which was trained on ChEMBL~\cite{gaulton_chembl_2012}. We note that for the Aripiprazole Similarity experiment, this enables direct comparison to Double Loop RL~\cite{bjerrum_faster_2023} as the authors also used the same Prior. The hyperparameters used for Experiment 1: Aripiprazole Similarity and Experiment 3: Dopamine Type 2 Receptor (DRD2) are presented in Table \ref{table: hyperparameters}.

\begin{table}[ht!]
    \centering
    \caption{Hyperparameters (default) used in Experiment 1: Aripiprazole Similarity and Experiment 3: Dopamine Type 2 Receptor (DRD2).}
    \begin{tabular}{|l|c|c|c|c|c|}
        \hline
        Algorithm & Sigma ($\sigma$) & Batch Size & Learning Rate & k & Alpha ($\alpha$) \\[2pt]
        \hline
        Augmented Memory & 128 & 64 & 0.0001 & N/A & N/A \\[2pt]
        \hline
        REINVENT~\cite{olivecrona_molecular_2017, blaschke_reinvent_2020} & 128 & 64 & 0.0001 & N/A & N/A \\[2pt]
        \hline
        Augmented Hill-Climbing (AHC)~\cite{thomas_augmented_2022} & 60 & 64 & 0.0001 & 0.5 & N/A \\[2pt]
        \hline
        Best Agent Reminder (BAR)~\cite{atance2022novo} & 1 & 64 & 0.0001 & N/A & 0.5 \\[2pt]
        \hline
    \end{tabular}
    \label{table: hyperparameters}
\end{table}

\section{Practical Molecular Optimization (PMO) Hyperparameters}
\label{appendix:pmo-tuning}

\begin{table}[ht]
    \centering
    \caption{Hyperparameters used in Experiment 2: Practical Molecular Optimization (PMO)~\cite{gao_sample_2022} Benchmark.}
    \begin{tabular}{|l|c|c|c|c|c|}
        \hline
        Algorithm & Sigma ($\sigma$) & Batch Size & Learning Rate & k & Alpha ($\alpha$) \\[2pt]
        \hline
        Augmented Memory & 500 & 64 & 0.0005 & N/A & N/A \\[2pt]
        \hline
        REINVENT~\cite{olivecrona_molecular_2017, blaschke_reinvent_2020} & 500 & 64 & 0.0005 & N/A & N/A \\[2pt]
        \hline
        Augmented Hill-Climbing (AHC)~\cite{thomas_augmented_2022} & 120 & 256 & 0.0005 & 0.25 & N/A \\[2pt]
        \hline
        Best Agent Reminder (BAR)~\cite{atance2022novo} & 1000 & 64 & 0.0005 & N/A & 0.25 \\[2pt]
        \hline
    \end{tabular}
    \label{table: pmo-hyperparameters}
\end{table}

\begin{table}[ht!]
    \centering
    \caption{Best Agent Reminder (BAR)~\cite{atance2022novo} hyperparameter tuning for Experiment 2: Practical Molecular Optimization (PMO)~\cite{gao_sample_2022}. The AUC top-10 was used to assess performance and was based on the protocol proposed in the PMO benchmark: average AUC top-10 across 3 independent runs of zaleplon\_mpo and perindopril\_mpo.}
    \begin{tabular}{|l|c|c|c|}
        \hline
        Sigma ($\sigma$) & Alpha ($\alpha$) & Top-10 AUC\\[2pt]
        \hline
        250 & 0.25 & 0.610 \\[2pt]
        \hline
        250 & 0.50 & 0.677 \\[2pt]
        \hline
        500 & 0.25 & 0.708 \\[2pt]
        \hline
        500 & 0.50 & 0.728 \\[2pt]
        \hline
        750 & 0.25 & 0.739 \\[2pt]
        \hline
        750 & 0.50 & 0.732 \\[2pt]
        \hline
        \textbf{1000} & \textbf{0.25} & \textbf{0.762} \\[2pt]
        \hline
        1000 & 0.50 & 0.759 \\[2pt]
        \hline
    \end{tabular}
    \label{table: bar-mpo-hyperparameter-tuning}
\end{table}

The hyperparameters used for the Practical Molecular Optimization (PMO)~\cite{gao_sample_2022} benchmark is presented in Table \ref{table: pmo-hyperparameters}. The hyperparameters provided in the PMO repository for REINVENT~\cite{olivecrona_molecular_2017, blaschke_reinvent_2020} and AHC~\cite{thomas_augmented_2022} were used. The hyperparameters for BAR~\cite{atance2022novo} were tuned according to Table \ref{table: bar-mpo-hyperparameter-tuning}. We note that the default $\sigma$ hyperparameter is 1 as stated in the BAR repository. However, we found that the resulting AUC Top-10 was much lower than all $\sigma$ values in \ref{table: bar-mpo-hyperparameter-tuning}. Thus, we performed hyperparameter tuning using much larger $\sigma$ values according to the values REINVENT was tuned with.

\section{Practical Molecular Optimization (PMO) Augmented Memory and BAR Prior}

\begin{table}[ht!]
    \centering
    \caption{LSTM model hyperparameters for Augmented Memory and BAR}
    \begin{tabular}{|l|c|c|}
        \hline
        Cell Type & LSTM \\[2pt]
        \hline
        Number of Layers & 3 \\[2pt]
        \hline
        Embedding Layer Size & 256 \\[2pt]
        \hline
        Dropout & 0 \\[2pt]
        \hline
        Training Batch Size & 128 \\[2pt]
        \hline
        SMILES Training Randomization  & True \\[2pt]
        \hline
    \end{tabular}
    \label{table: pmo-prior-training}
\end{table}

Augmented Memory required training a Prior and follows the protocol from REINVENT~\cite{blaschke_reinvent_2020} and using the provided ZINC~\cite{irwin_zinc20free_2020} dataset in the PMO~\cite{gao_sample_2022} repository. Table \ref{table: pmo-prior-training} shows the hyperparameters of the LSTM~\cite{hochreiter1997long} network. We note all hyperparameters were kept default and the model was trained for 10 epochs as SMILES validity reached 95\% and the total wall time was 11 minutes 57 seconds. BAR~\cite{atance2022novo} experiments were run with this same pre-trained Prior.

\section{DRD2 Experiment Wall Times}

\begin{table}[ht!]
    \centering
    \caption{Experiment 3: Dopamine Type 2 Receptor (DRD2) Case Study wall times.}
    \begin{tabular}{|l|c|c|}
        \hline
        Algorithm & Wall Time \\[2pt]
        \hline
        Augmented Memory & 21 hours 25 minutes $\pm$ 2 hours 20 minutes \\[2pt]
        \hline
        REINVENT & 27 hours 7 minutes $\pm$ 2 hours 21 minutes \\[2pt]
        \hline
        Augmented Hill-Climbing (AHC) & 32 hours 25 minutes $\pm$ 5 hours 38 minutes \\[2pt]
        \hline
        Best Agent Reminder (BAR) & 37 hours 42 minutes $\pm$ 4 hours 39 minutes \\[2pt]
        \hline
    \end{tabular}
    \label{table: drd2-wall-times}
\end{table}

The wall times for Experiment 3: Dopamine Type 2 Receptor (DRD2) are presented in Table \ref{table: drd2-wall-times}. We note that we performed a total of 6 replicates for each algorithm: 3 in the main result experiments and 3 in the exploiting AutoDock Vina experiments (Figure \ref{fig:appendix-exploit-vina}). For REINVENT~\cite{olivecrona_molecular_2017, blaschke_reinvent_2020}, AHC~\cite{thomas_augmented_2022}, and BAR~\cite{atance2022novo}, we pool the experiments using experience replay. For example, REINVENT values are reported based on 12 total runs: 3 for main result experiments, 3 for main results experiments with experience replay, 3 for exploiting AutoDock Vina experiments, and 3 for exploiting AutoDock Vina experiments with experience replay.  The bottleneck in all experiments is AutoDock Vina~\cite{trott2010autodock} and the wall time is highly variable, depending on the molecules sampled by the Agent. Finally, we note that all experiments were run with a batch size of 64 for 150 epochs. The exception is BAR which was run for 75 epochs as each epoch samples 2 batches of molecules: one from the current Agent and one from the best Agent. All experiments had an AutoDock Vina oracle budget of 9,600 calls. Finally, we comment on the variable wall times of each algorithm despite having a fixed oracle budget. There are two sources of stochasticity. Firstly, the experiments were performed on a shared cluster and compute speed is variable depending on usage. Secondly, docking is itself stochastic and generally requires more search time for larger molecules. Augmented Memory jointly optimizes for Vina, QED~\cite{bickerton_quantifying_2012}, and MW which generally enforces smaller molecules and could be a reason for the faster average compute time.

\section{AutoDock Vina DRD2 Receptor Preparation and Docking}

The receptor grid for AutoDock Vina~\cite{trott2010autodock} docking against DRD2 (PDB ID: 6CM4~\cite{wang2018structure}) was performed using DockStream~\cite{guo_dockstream_2021}. The PDB file for 6CM4 was first downloaded from the Protein Data Bank. One monomer unit was extracted and refined using PDBFixer~\cite{eastman2017openmm} through the DockStream wrapper. The prepared grid was centered at (x, y, z) = (9.93, 5.85, -9.58) with a search box of 15Å x 15Å x 15Å. Docking for all experiments were performed with DockStream using the following protocol: embed sampled SMILES with RDKit Universal Force Field (UFF)~\cite{rappe1992uff} with 600 maximum convergence iterations and execute AutoDock Vina docking parallelized over 36 CPU cores (Intel(R) Xeon(R) Platinum 8360Y processors).

\section{Proof of Loss Function and Policy Gradient Equivalency}
\label{loss-proof}
In this section, we show that the loss function used to tune the Agent is equivalent to optimizing the expected reward of the policy following the REINFORCE~\cite{williams_simple_1992} algorithm. Molecules are represented as a sequence of tokens given by the Simplified Molecular Input Line Entry System (SMILES)~\cite{weininger_smiles_1988} format and generated in an autoregressive manner. The generative process is Markovian (Equation \ref{eq:appendix-markovian}):

\begin{equation}
P(x) = \prod_{t=1}^{T} P(s_t \mid s_{t-1}, s_{t-2}, \ldots, s_1)
\label{eq:appendix-markovian}
\end{equation}

Equation 1 states that the probability of generating a given SMILES, $x$, is equal to the product of the probabilities of generating a token at time-step $t$, given the sequence so far at time-step $t-1$. The model is pre-trained on a dataset of molecules (ChEMBL~\cite{gaulton_chembl_2012} for the main experiments and ZINC~\cite{irwin_zinc20free_2020} for the benchmarking experiment) to yield the Prior which is parameterized by the weights $\theta$. The Agent is initialized identical to the Prior but is fine-tuned during the reinforcement learning (RL) process. The Augmented Likelihood is defined as a linear combination between the Prior and a reward term (Equation \ref{eq:appendix-augmented-likelihood}):

\begin{equation}
\log\pi_{\theta_{\textsubscript{Augmented}}} = \log\pi_{\theta_{\textsubscript{Prior}}} + \sigma S(x)
\label{eq:appendix-augmented-likelihood}
\end{equation}

$S$ is the reward function assessing the desirability of a sampled molecule and $\sigma$ is a hyperparameter that scales the reward. A higher $\sigma$ places a greater contribution on the reward function and less on the Prior. The Prior is used to ensure generated SMILES are syntactically correct and has been empirically shown to enforce reasonable chemistry. The loss function is defined as the squared difference between the Augmented Likelihood and the Agent Likelihood for a given batch, $B$, of sampled SMILES constructed following the actions, $ a \in A^* $ (Equation \ref{eq:appendix-dap}):

\begin{equation}
L(\theta) = \frac{1}{|B|} \left[\sum_{a \in A^*}(\log\pi_{\theta_\textsubscript{Augmented}} - \log\pi_{\theta_\textsubscript{Agent}})\right]^2
\label{eq:appendix-dap}
\end{equation}

Taking the derivative with respect to $\theta$ (Equation \ref{eq:appendix-dap-derivative}):

\begin{equation}
\nabla_{\theta} L(\theta) = -2 \frac{1}{|B|} \left[ \sum_{a \in A^*} \log \pi_{\theta_{\textsubscript{Augmented}}} - \log \pi_{\theta_{\textsubscript{Agent}}} \right] \sum_{a \in A^*} \nabla_{\theta} \log \pi_{\theta_{\textsubscript{Agent}}}
\label{eq:appendix-dap-derivative}
\end{equation}

Minimizing $J(\theta)$ tunes the Agent to generate molecules satisfying the reward function. \\

Following Fialková et al.~\cite{fialkova_libinvent_2022}, we now show that minimizing $J(\theta)$ is equivalent to optimizing the expected reward of the policy. The generative process is cast as an on-policy RL problem by defining the state space, $S_t$, and the action space, $A_t(s_t)$. $S_t$ denotes every intermediate sequence of tokens leading up to the fully constructed SMILES and $A_t(s_t)$ are the token sampling probabilities at every intermediate state. $A_t(s_t)$ is controlled by the policy, $\pi_{\theta}$, which is parameterized by the weights, $\theta$, of the neural network. Given a reward function, $R$, the objective is to maximize the expected reward when taking actions defined by the policy (Equation \ref{eq:appendix-expected-reward}):

\begin{equation}
J(\theta) = \mathbb{E}_{a_t \sim \pi_\theta} \left[ \sum_{t=0}^{T} R(a_t, s_t) \right]
\label{eq:appendix-expected-reward}
\end{equation}

Rewriting the expectation (Equation \ref{eq:appendix-double-sum}):

\begin{equation}
J(\theta) = \sum_{t=0}^{T} \sum_{a \in A_t} R(a_t, s_t) \pi_{\theta}(a_t | s_t)
\label{eq:appendix-double-sum}
\end{equation}

The expectation can be rewritten as a double summation over all time-steps and actions taken at each time-step, following the policy, $\pi_{\theta}$. Next, the derivative of the expression is taken (Equation \ref{eq:appendix-double-sum-derivative}):

\begin{equation}
\nabla_{\theta} J(\theta) = \sum_{t=0}^{T} \sum_{a \in A_t} R(a_t, s_t) \nabla_{\theta} \pi_{\theta}(a_t | s_t)
\label{eq:appendix-double-sum-derivative}
\end{equation}

Applying the log-derivative trick (Equation \ref{eq:appendix-log-derivative-trick}):

\begin{equation}
\nabla_{\theta} J(\theta) = \sum_{t=0}^{T} \sum_{a \in A_t} R(a_t, s_t) \pi_{\theta}(a_t | s_t) \nabla_{\theta} \log \pi_{\theta}(a_t | s_t)
\label{eq:appendix-log-derivative-trick}
\end{equation}

Using the definition of expectation for discrete variables, i.e., the policy actions which can only sample the vocabulary tokens (Equation \ref{eq:appendix-discrete-expectation}):

\begin{equation}
\nabla_{\theta} J(\theta) = \mathbb{E}_{a_t \sim \pi_{\theta}} \left[ \sum_{t=0}^{T} R(a_t, s_t) \nabla_{\theta} \log \pi_{\theta}(a_t | s_t) \right]
\label{eq:appendix-discrete-expectation}
\end{equation}

As computing the expectation is intractable, it is instead approximated by sampling a batch, $B$, of trajectories, i.e., SMILES strings. The process of SMILES generation is further defined as an episodic task where reward is only given at the terminal state. In particular, the \textit{desirability} of a SMILES sequence only applies once the full SMILES string has been sampled and it maps to a valid molecule. Thus, all intermediate rewards are 0. Defining the set of actions taken in a batch, $A^*$ as the specific token sequences generated at a given epoch yields Equation \ref{eq:appendix-terminal-batch}:

\begin{equation}
\nabla_{\theta} J(\theta) = \frac{1}{|B|} \left[ \sum_{t=0}^{T} \sum_{a \in A^*} R(a_t, s_t) \nabla_{\theta} \log \pi_{\theta}(a_t | s_t) \right]
\label{eq:appendix-terminal-batch}
\end{equation}

Finally, the reward, $R$ is defined according to Fialková et al.~\cite{fialkova_libinvent_2022} (Equation \ref{eq:appendix-reward-choice}):

\begin{equation}
R(a_t, s_t) = \log \pi_{\theta_{\textsubscript{Augmented}}} - \log \pi_{\theta_{\textsubscript{Agent}}}
\label{eq:appendix-reward-choice}
\end{equation}

Substituting Equation \ref{eq:appendix-reward-choice} into Equation \ref{eq:appendix-terminal-batch} yields the desired equivalency to the loss function (Equation \ref{eq:appendix-dap-derivative}) up to a constant factor.

\section{Augmented Memory Algorithm}
\label{algorithm}

The pseudo-code for Augmented Memory is presented here.

\begin{algorithm}[H]
    \caption{Augmented Memory}
    \SetKwInput{Input}{Input}
    \SetKwInput{Output}{Output}
    \Input{Prior $\pi_\textsubscript{Prior}$, Epochs $N$, Augmentation Rounds $A$, Scoring Function $S$, Sigma $\sigma$}
    \Output{Fine-tuned Agent Policy $\pi_{\theta_{\textsubscript{Agent}}}$, Generated Molecules $G$}
    \textbf{Initialization:}
    
    Generative Agent $\pi_{\theta_{\textsubscript{Agent}}}$ = $\pi_\textsubscript{Prior}$;
    
    Diversity Filter $DF$;
    
    Replay Buffer $B = \{\}$;

    \For{$i \leftarrow 1$ \KwTo $N$} {
        Sample batch of SMILES $X = \{ x_1, \ldots , x_b \}$ with $x_i \sim \pi_{\theta_{\textsubscript{Agent}}}$;

        Compute reward using the scoring function $ S(X) $;\\[3pt]

        Modify reward based on the diversity filter $ DF (S(X))$;\\[3pt]

        Update replay buffer $B_i = X_{i} \cup X_{i-1}$;\\[3pt]

        (Optionally) purge replay buffer;\\[3pt]
        
        Compute Augmented Likelihood $\log\pi_{\theta_{\textsubscript{Augmented}}} = \log\pi_{\textsubscript{Prior}}(X) + \sigma S(X)$;\\[3pt]

        Compute loss $J(\theta) = (\log\pi_{\textsubscript{Augmented}} - \log\pi_{\theta_\textsubscript{Agent}}(X))^2$;\\[3pt]

        Update the Agent's policy $\pi_{\theta_{\textsubscript{Agent}}}$;

        \For{$j \leftarrow 1$ \KwTo $A$} {
        Augment sampled SMILES $X_\textsubscript{Augmented}$;\\[3pt]

        Compute Augmented Likelihood of augmented SMILES (reward is unchanged) $\log\pi_{{\textsubscript{Augmented}}} = \log\pi_{\textsubscript{Prior}}(X_\textsubscript{Augmented}) + \sigma S(X)$;\\[3pt]

        Compute loss $J(\theta)_{\textsubscript{Augmented}} = (\log\pi_{\textsubscript{Augmented}} - \log\pi_{\theta_\textsubscript{Agent}}(X_\textsubscript{Augmented}))^2$;\\[3pt]

        Augment entire replay buffer $B_\textsubscript{Augmented}$;\\[3pt]

        Compute Augmented Likelihood on the augmented buffer (reward is the buffer stored rewards) $\log\pi_\textsubscript{Buffer Augmented} = \log\pi_{\textsubscript{Prior}}(B_\textsubscript{Augmented}) + \sigma S(B)$;\\[3pt]

        Compute augmented buffer loss $J(\theta)_{\textsubscript{Buffer Augmented}} = (\log\pi_{\textsubscript{Buffer Augmented}} - \log\pi_{\theta_\textsubscript{Agent}}(B_\textsubscript{Augmented}))^2$;\\[3pt]

        Concatenate the augmented sampled SMILES loss and the augmented buffer loss $J(\theta){\textsubscript{Augmented Memory}} = J(\theta)_{\textsubscript{Augmented}} + J(\theta)_{\textsubscript{Buffer Augmented}}$;\\[3pt]

        Update the Agent's policy $\pi_{\theta_{\textsubscript{Agent}}}$;
        
    }
}
    
\end{algorithm}
\end{document}